\documentclass[usegraphicx,usenatbib,useAMS]{mn2e}
\usepackage{times}
\usepackage{epsfig}

\begin{document}

\newcommand{\be}{\begin{equation}}
\newcommand{\ee}{\end{equation}}
\newcommand{\ba}{\begin{eqnarray}}
\newcommand{\ea}{\end{eqnarray}}
\newcommand{\mat}{{\bf}}
\newcommand{\nn}{\nonumber\\}
\newcommand{\bA}{{\bf A}}
\newcommand{\bE}{{\bf E}}
\newcommand{\bB}{{\bf B}}
\newcommand{\bC}{{\bf C}}
\newcommand{\bF}{{\bf F}}
\newcommand{\bS}{{\bf S}}
\newcommand{\bfa}{{\bf a}}
\newcommand{\bb}{{\bf b}}
\newcommand{\bg}{{\bf g}}
\newcommand{\bj}{{\bf j}}
\newcommand{\bk}{{\bf k}}
\newcommand{\bn}{{\bf n}}
\newcommand{\bp}{{\bf p}}
\newcommand{\br}{{\bf r}}
\newcommand{\bu}{{\bf u}}
\newcommand{\bv}{{\bf v}}
\newcommand{\bx}{{\bf x}}
\newcommand{\by}{{\bf y}}
\newcommand{\bbmu}{\mbox{\boldmath $\mu$}}
\newcommand{\dd}{{\partial}}
\newcommand{\ddt}{{\partial\over \partial t}}
\newcommand{\lnL}{\ln{\cal L}}
\newcommand{\bbtheta}{\mbox{\boldmath $\theta$}}
\def\gs{\mathrel{\raise1.16pt\hbox{$>$}\kern-7.0pt 
\lower3.06pt\hbox{{$\scriptstyle \sim$}}}}         
\def\ls{\mathrel{\raise1.16pt\hbox{$<$}\kern-7.0pt 
\lower3.06pt\hbox{{$\scriptstyle \sim$}}}}         

\title[The Mass Function of the Stellar Component of SDSS
Galaxies]{The Mass Function of the Stellar Component of Galaxies
in the
  Sloan Digital Sky Survey}
\author[Panter, Heavens \& Jimenez]
{Benjamin Panter$^1$\thanks{email: bdp@roe.ac.uk; afh@roe.ac.uk;
    raulj@physics.upenn.edu},
Alan F. Heavens$^1$\footnotemark[1], Raul Jimenez$^2$\footnotemark[1] \\
$^1$Institute for Astronomy, University of Edinburgh, Blackford
Hill,
Edinburgh EH9 3HJ, UK\\
$^2$Department of Physics and Astronomy, University of
Pennsylvannia, 209 South 33rd Street, Philadelphia, Pennsylvania,
19104-6396, USA}

\maketitle

\begin{abstract}
  Using the MOPED algorithm we determine non-parametrically the
  Stellar Mass Function of 96,545 galaxies from the Sloan Digital Sky
  Survey data release one. By using the reconstructed spectrum due to
  starlight we can eliminate contamination from either emission lines
  or AGN components. Our results give excellent agreement with
  previous works, but extend their range by more than two decades in
  mass to $10^{7.5} \ls M_s/h^{-2}M_\odot \ls 10^{12}$.  We present both a
  standard Schechter fit and a fit modified to include an extra,
  high-mass contribution, possibly from cluster cD galaxies. The
  Schechter fit parameters are $\phi^*=(7.7\pm 0.8)\times 10^{-3} h^3
  Mpc^{-3}$, $M^*=(7.53 \pm 0.04) \times 10^{10} h^{-2}M_\odot$ and
  $\alpha=-1.167\pm 0.004$.  Our sample also yields an estimate for
  the contribution from baryons in stars to the critical density of
  $\Omega_{b*}h=(2.40 \pm 0.04) \times 10^{-3}$, in good agreement
  with other indicators.  Error bars are statistical and a Salpeter
  IMF is assumed throughout. We find no evolution of the mass function
  in the redshift range $0.05 < z < 0.34$ indicating that almost all
  stars were already formed at $z \sim 0.34$ with little or no star
  formation activity since then and that the evolution seen in the
  luminosity function must be largely due to stellar fading.
\end{abstract}


\section{Introduction}
Recent advances in galaxy modelling codes mean that we are on the
verge of being able to predict reliably the mass function of
stellar component of galaxies in the Universe. To match this
modelling we require a good observational determination from a
large sample of galaxies. This has been attempted in the past by
combining the Two Micron All Sky Survey (2MASS) with the 2dF
Galaxy Redshift Survey (2dF GRS) by \cite{Cole01} and the Sloan
Digital Sky Survey (SDSS) Early Data Release (EDR) by
\cite{Bell}. Both these methods used the K-band magnitudes of
galaxies in the 2MASS survey to estimate the mass of their
respective redshift survey galaxies.

Unfortunately neither of the redshift surveys can be used to their
full depth when combined with the 2MASS survey - it does not go deep
enough. The Sloan Digital Sky Survey (SDSS) Data Release One (DR1,
\cite{DR1}, \cite{Strauss}) gives a larger sample of galaxies (of the
order 100,000 spectra) and it would be advantageous to be able to form
a mass function from all of these galaxies. Obviously, just using the
optical magnitudes from these surveys is insufficient to yield stellar
mass, as some account must be taken of AGN component, dust component,
hot gas and the evolution of stellar population. It would be highly
unsatisfactory to estimate a blanket correction for these components,
and hence an independent determination of the stellar mass is required
for each individual galaxy in the survey.

With the advent of efficient spectral fitting algorithms it is
possible to extract and model the stellar populations of galaxies,
and reliably exclude the contribution to the spectrum of emission
lines, dust and AGN. The stellar component of the mass of the
galaxy can then be calculated, and when applied to a large enough
survey of galaxies be used to calculate the mass function of the
stellar component.

\begin{figure*}
\includegraphics[width=.7\textwidth]{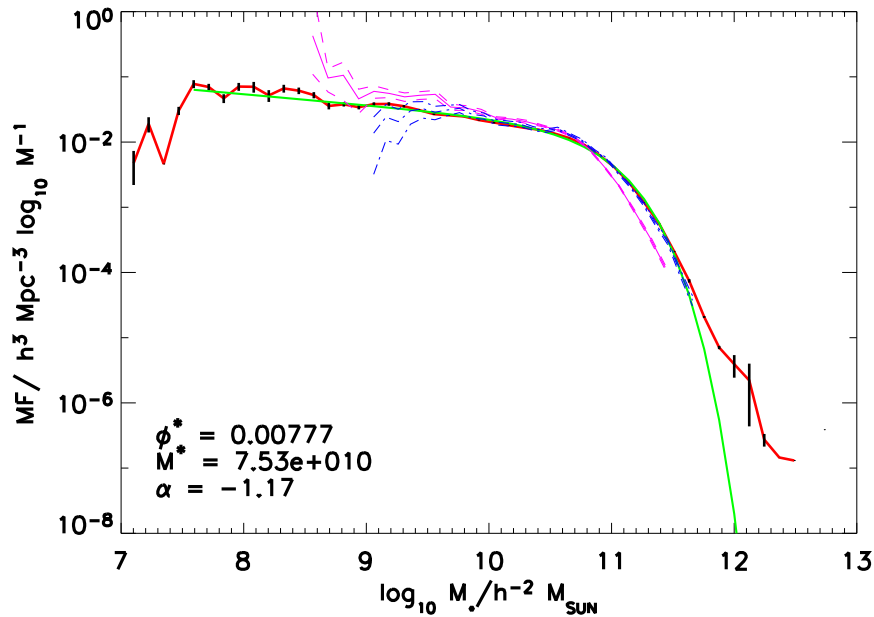}\\
\caption{Mass function for the SDSS DR1, with a Schechter Function
fit overplotted (solid green line). Also shown are the mass functions recovered by
Cole et al. (2001) (dashed blue line) and Bell et al. (2003) (solid
cyan line). The mass offset of the
Bell data is due to a different choice of IMF. The first three
bins have been excluded from the fit.}\label{mf_s}
\end{figure*}

\section{Deriving Stellar Masses with MOPED}
\subsection{MOPED}

The MOPED technique (\cite{HJL00}, \cite{Reichardt01}, \cite{Panter03},
\cite{nat04}) was used to extract the Star Formation History (SFH) of
the galaxies contained in the SDSS Main Galaxy catalogue
(\cite{Shen03}). The technique is examined in detail in previous
papers, and is equivalent to fitting synthetic stellar population
spectra to each galaxy spectrum using a novel data compression and
analysis algorithm.  The main emission-line regions are
excluded. Masses were calculated for a complete sample inside a
redshift range of $0.005 < z < 0.34$ with R band apparent magnitudes
$15.0 \leq m_R \leq 17.77$. Following the suggestion of \cite{Shen03}
we have also set a surface brightness limit of $\mu_R <
23.0$. Throughout we assume the concordant WMAP cosmology,
$\Omega_m=0.27$, $\Omega_v=0.73$, $H_0\equiv 100h \textrm{ km
s}^{-1} \textrm{Mpc}^{-1}$, $h=0.71$ (\cite{Spergel03}).

\subsection{Mass extraction}

MOPED extracts the star formation and metalicity history of a
galaxy along with its dust content. The star formation history of
each galaxy is modelled by 23 numbers: the star formation fraction
in each of 11 time periods, largely spaced equally in log(lookback
time); 11 associated metallicities of the star-forming gas; a
simple dust screen characterised by an LMC extinction curve (\cite{Gordon03}) and a
single dust parameter.  By combining this information with the
input stellar model spectra we can create a synthetic spectrum of
the stellar component of the galaxy spectrum.  MOPED adjusts the
23 parameters until the best match with 23 numbers derived from
the galaxy spectrum is found.  These parameters, combined with an
overall normalisation of the galaxy spectrum, allow the integrated
mass created in stars to be estimated.

Following \cite{Cole01} to account for the mass of stars lost
through winds and supernovae we use the recycling fraction
$R=0.28$ derived from stellar evolution theory. The final stellar
mass of the galaxy is then $(1-R)$ times the integrated mass of
the different fitted populations totalled over the bins.

The masses recovered by this method are dependent on choice of IMF. We
have chosen a Salpeter IMF to allow direct comparison with theory and
other predictions. Since we are using a Salpeter IMF with a low mass
cut off at $0.1$ M$_{\odot}$ we are insensitive to any mass
included in brown dwarfs.

\begin{figure*}
\includegraphics[width=.7\textwidth]{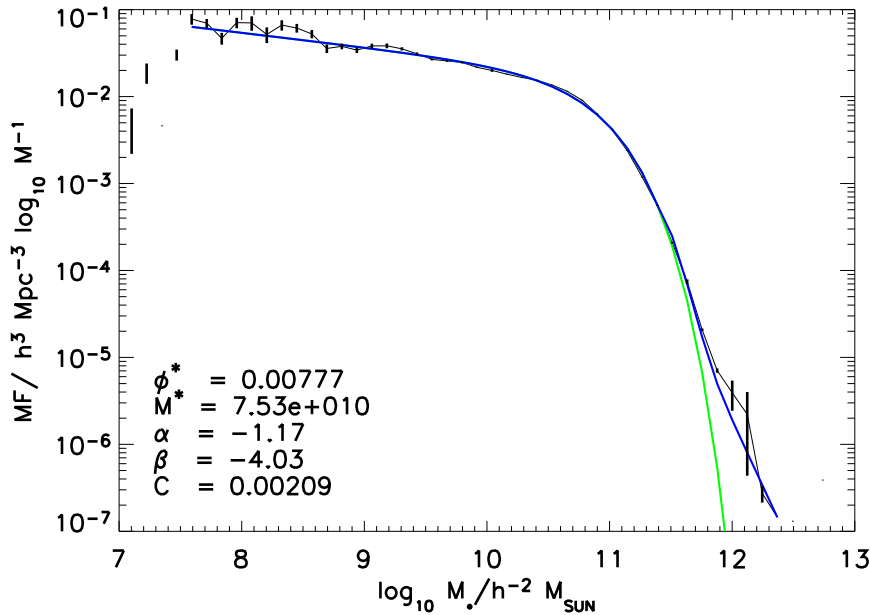}\\
\caption{Mass Function with Modified Fit. Here we present a
  well-fitting modification to the Schechter function which gives
    excellent agreement with our results at high mass. Again, the
    first three bins have been excluded from the fit. The Schechter function alone is a good fit up to $10^{11.5}h^{-2}M_\odot$}\label{mf_c}
\end{figure*}

To turn the measurements of mass to a mass density we weight each
galaxy by $1/V_{max}$, where $V_{max}$ is the maximum volume in
which a given galaxy would still be included within the limits of
the survey. In the past these have been calculated by assuming
some luminosity evolution and calculating a K-correction from
reference spectra. Since we have knowledge of the star formation
history of every galaxy in the SDSS we can evolve the luminosity
and surface brightness history of each galaxy rather than assume a
fixed evolution. To evolve the surface brightness of the galaxy we
assume that over the redshift range studied there has been no size
change of the galaxy.  The mass estimator should be unbiased; in
principle it is sensitive to clustering, but the effective volumes
probed are large enough that this should be a small effect.

To account for the three-arcsec fibre diameter we scale our masses
up by the ratio of the flux from the photometric $R$ band
petrosian magnitudes to that from the spectroscopically determined
fibre magnitudes. This is unlikely to succeed for individual
galaxies, but there is evidence from the Petrosian and fibre
colours that there is no systematic offset caused overall
(\cite{GB03}).  We also find that, for galaxies of about the same
mass, the intermediate results such as star formation fraction
show no signs of a trend with galaxy redshift, which one would
expect if there were a significant systematic error arising from
aperture effects (\cite{BCTpaper}).

To calculate the statistical errors on our mass function we
applied a bootstrap error algorithm. The binning operation was
performed with 1000 randomly-selected resamples of the original
data set, and the standard deviations of the heights of each peak
recovered.

\section{Results}

\subsection{Galaxy Stellar Mass Function}

The stellar mass function is shown in Fig. \ref{mf_s}. Between
about $10^9$ and $10^{11} h^{-2} M_\odot$ we find excellent agreement
with results obtained by previous studies of SDSS and 2dFGRS
galaxies (\cite{Bell} and \cite{Cole01}), where the stellar mass is
estimated more simply from infrared data.  We are able to extend
the mass range considerably, by around a decade in mass at the
upper mass end, and about two decades at the lower-mass end. The
stellar mass function of SDSS galaxies is now accurately
determined between $10^{7.5}$ and $10^{12}\ h^{-2}M_\odot$, where
$h$ is the Hubble parameter in units of $100
km\,s^{-1}\,Mpc^{-1}$.

We fit the galaxy stellar mass function with a Schechter function
(\cite{Schechter})
\begin{equation}
\phi(M_s) dM_s =
\phi^*\left(\frac{M_s}{M^*}\right)^\alpha\exp\left(\frac{-M_s}{M^*}\right)dM_s
\end{equation}
with best-fitting parameters $\phi^*=(7.7\pm 0.8)\times 10^{-3}
h^{3}Mpc^{-3}$, $\alpha=-1.167\pm 0.004$ and $M^*=(7.53\pm 0.04)
\times 10^{10} h^{-2}M_\odot$. This fit is shown overplotted in
Fig.\ref{mf_c}, and is a good fit up to $M_s=10^{11.5} h^{-2}M_\odot$.

There is evidence for an excess over the Schechter fit at the
high-mass end (which seems to be confirmed from dynamical measurements
of the mass of SDDS galaxies, Bernardi et al. private communication),
which can be well modelled by the addition of a power law over
the range $11.5 < \log_{10}(h^2 M_s/M_\odot) < 12.6$ of the form
\begin{equation}
\phi_c=\phi+F_C\left(\frac{M_s}{M^*}\right)^\beta
\end{equation}
with $\beta=-4.03 \pm 0.03$ and $F_C=(2.1 \pm 0.2) \times 10^{-3}
h^3Mpc^{-3}$ as shown in Figure \ref{mf_c}. This excess could be due to cD
galaxies or a failure of the modelled correction to total magnitudes for these
extremely large galaxies.

\begin{figure*}
  \includegraphics[width=0.7\textwidth]{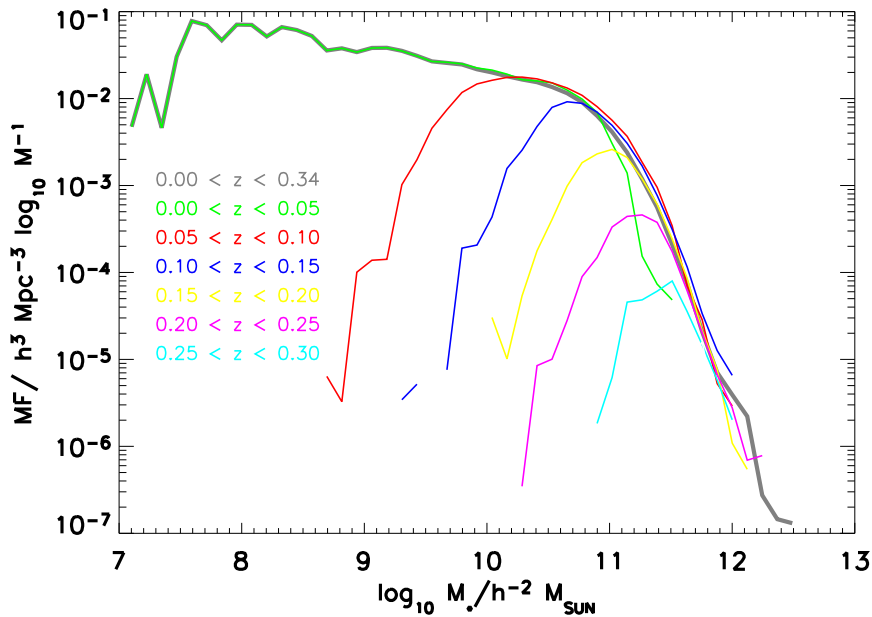}\\
  \caption{Mass functions for different redshift ranges. The agreement is 
    generally very good where the samples overlap, indicating that
    there is little evolution over the redshift range $0-0.3$.  There
    is some discrepancy at the high-mass end in the lowest redshift
    range, thought to be due to Sloan photometric pipeline shredding large galaxy
    images.}\label{mf_z}
\end{figure*}

\subsection{Evolution with Redshift}

By splitting the DR1 sample by redshift we can probe the evolution
of the stellar mass function in recent times. Fig. \ref{mf_z}
shows the stellar mass function for galaxies within narrow
redshift ranges. Because of the flux limit, there is essentially a
minimum mass which can be probed at each redshift, but this is not
a sharp cutoff because the galaxies have a range of star formation
histories so the mapping from stellar mass to luminosity is not
one-to-one. It is apparent from the figure that within the limits
of the survey there is very little, if any, evolution in the
redshift range $0.05 < z < 0.34$.  The only notable deviation from
this is an apparent deficiency of high-mass galaxies
($M_s>10^{11}h^{-2}M_\odot$) at very low $z\sim 0.05$.
The high mass results from the lowest redshift sample should be
treated with caution. The galaxies at the high mass end of the
mass function are generally large in their angular size. This
leads to a problem with ``shredding'' by the SDSS photometric
pipeline, where large galaxies are treated as many smaller
sources. It is thought that this is only really a problem for
$z<0.02$, but for $z<0.01$ as many as 10\% of the detections could
be affected (SDSS Collaboration, private communication). 

\begin{figure}
  \includegraphics[width=0.5\textwidth]{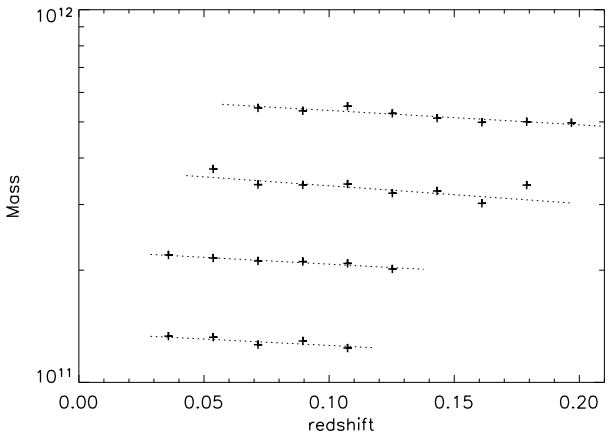}\\
  \caption{The evolution of the average galaxy stellar mass for galaxies with
  (from the bottom) $\log_{10}(L_R/L_\odot) = 10.25-10.3, 10.45-10.5, 10.65-10.7, 10.85-10.9$, where $L_R$ is the
  $R-$band luminosity, $K-$corrected using kcorrect v3\_1B (Blanton et
  al (2003a))}
  \label{evol}
\end{figure}

The lack of evolution of the mass function with redshift is in
contrast to the significant evolution found in the luminosity
function, where the characteristic luminosity has become fainter
by around 0.3 magnitudes since $z=0.2$, and the number density of
bright galaxies has declined by a factor of two or more
(\cite{Loveday,Blanton2003}).  The most natural explanation is
that the stellar mass content has hardly changed, but that the
galaxies have just become significantly fainter;  this is expected
given the drop-off in star formation activity, and can be
illustrated by Fig. \ref{evol}, which shows the evolution of the
average stellar mass with redshift, for galaxy populations of
different luminosities.  We see clearly an increase in the average
mass with decreasing redshift.

\subsection{Cosmological Baryon density in stars}
Our results can be used to give a further constraint on
the contribution to the density parameter from baryons in stars, 
$\Omega_{b*}$. By integrating the mass over the range of the
mass function we deduce a value of $\Omega_{b*}h=2.40\pm
0.04 \times 10^{-3}$. This value is in broad agreement with
results obtained previously
(\cite{Cole01,Bell,fukugita,kochanek,GB03}).
Our error is a bootstrap estimate, and is purely statistical;
systematic errors such as the choice of IMF have not been included.

\section{Discussion and Conclusions} 
We have calculated the stellar mass function
of 96,545 galaxies in the Data Release of the Sloan Digital Sky
Survey over  about 5 decades of mass. The results are in good
agreement with previous studies, where the stellar masses were
estimated more simply from infrared data.  The range probed is
considerably extended, and differences in derived parameters are
most likely due to different assumed initial mass functions.

Also of importance is the difference in numerical size and
redshift range of the sample used to generate the stellar mass
function, since we are not restricted to galaxies which appear in
the range of galaxies in 2MASS.

In contrast to the luminosity function, the mass function shows no
evidence for evolution with redshift.  The luminosity function
shows a fading of the characteristic luminosity, by a factor of
about 1.35 since $z=0.2$ (\cite{Loveday,Blanton2003}).  The
simplest interpretation is that the galaxy stellar masses do not
evolve significantly (and this is supported by the star formation
rates reported by \cite{nat04}), but that individual
galaxies fade.  Fig. \ref{evol} shows the average mass of massive
galaxies in small luminosity redshift ranges, as a function of the
observed redshift.  We see the expected trend:  massive galaxies
are typically brighter at higher redshift, scaling roughly as
$\log_{10}M_s(z) \simeq \log_{10}M_s(z=0)-0.4 z$.  This represents
a typical fading of around 20\% over the redshift range $z=0.2$ to
the present.

\section*{acknowledgments}

The authors wish to thank Jim Dunlop and John Peacock for useful
discussions contributing to the development of this work.

Funding for the creation and distribution of the SDSS Archive has been
provided by the Alfred P. Sloan Foundation, the Participating Institutions,
the National Aeronautics and Space Administration, the National Science
Foundation, the U.S.  Department of Energy, the Japanese Monbukagakusho, and
the Max Planck Society. The SDSS Web site is {\tt http://www.sdss.org/}.

The SDSS is managed by the Astrophysical Research Consortium (ARC) for the
Participating Institutions. The Participating Institutions are The University
of Chicago, Fermilab, the Institute for Advanced Study, the Japan
Participation Group, The Johns Hopkins University, Los Alamos National
Laboratory, the Max-Planck-Institute for Astronomy (MPIA), the
Max-Planck-Institute for Astrophysics (MPA), New Mexico State University,
University of Pittsburgh, Princeton University, the United States Naval
Observatory and the University of Washington.


\end{document}